\documentclass[aps,pra,twocolumn,showpacs]{revtex4}

\usepackage{bm}
\usepackage{epsfig}

\newcommand{\p}{\partial}
\newcommand{\half}{\frac{1}{2}}
\newcommand{\eab}{\epsilon_{\alpha\beta}}
\newcommand{\bnabla}{\bm{\nabla}}

\newcommand{\pontryagin}{{\cal N}}

\newcommand{\lex}{\ell_{\rm ex}}

\newcommand{\qfactor}{q}

\begin{document}

\title{Transmutation of momentum into position in magnetic vortices}
\author{Stavros Komineas$^1$ and Nikos Papanicolaou$^2$}
\affiliation{$^1$Max-Planck Institute for the Physics of Complex Systems,
N\"othnitzer Str. 38, 01187 Dresden, Germany \\
$^2$Department of Physics, and Institute of Plasma Physics, University of Crete,
Heraklion, Greece}

\date{\today}

\begin{abstract}
We show that transmutation of linear momentum into position
may occur in a system of three magnetic vortices thanks to a direct link between topology and dynamics in a ferromagnet.
This happens via exchange between the linear momentum of a vortex-antivortex
pair and the position of a single vortex during a semi-elastic
scattering process.
Vortex polarity switching occurs in the case of inelastic collisions.
\end{abstract}

\maketitle

\section{Introduction}
\label{sec:intro}

Magnetic vortices have been discussed traditionally in the context of
two-dimensional ferromagnets \cite{hubert,huber82,mertens99}.
They are magnetization configurations with a nontrivial topological
structure and are thus characterized by a topological invariant.
Interest in vortices has increased dramatically in recent years
due to many related observations in mesoscopic ferromagnetic elements,
such as disc-shaped thin elements with submicron 
dimensions \cite{raabe00,shinjo00,wachowiak02} where vortices are created spontaneously.
The magnetisation vector is tangent to the side boundary of the disc
and points along the disc axis at the disc center.
Similarly,
an antivortex is a configuration where the magnetisation vector winds around
the center in a sense opposite to that of a vortex
and has been observed in a specially designed magnetic element
\cite{shigeto02}, and as a metastable state in a ring element \cite{castano03}.
A significant feature of a vortex or antivortex is that its energy increases  logarithmically
with the size of the element and would thus tend to infinity on an infinite
film. This fact may explain why experimental observation of a single vortex is
relatively recent.

It is also possible to construct nontrivial magnetic configurations in
the form of vortex-antivortex (VA) pairs. Unlike single vortices, VA pairs are
localised configurations
and have finite energy even on an infinite film.
Vortex-antivortex pairs have been predicted
to be created by alternating external fields in an early study based on collective coordinates
\cite{pokrovskii85}.
They play a central role in dynamical processes
that lead to vortex polarity switching in ferromagnetic elements
as reported in recent experiments \cite{neudert05,waeyenberge06,yamada07}.

A subtle feature of a vortex is its {\it polarity}, i.e., the direction of the
magnetisation at the vortex center, usually called the vortex core.
Therefore, in a VA pair the vortex and the antivortex may have either the same  or opposite polarities.
In the case of like polarities, a VA pair undergoes Kelvin motion in a direction
perpendicular to the line connecting the vortex and the antivortex
\cite{papanicolaou99}, with velocity that is inversely proportional
to the distance between the two vortex centers.
Such a VA pair will be called  a Kelvin pair.
In the case of opposite polarities, a VA pair undergoes rotational motion around
a fixed guiding center,
with angular frequency that is inversely proportional to the square of the
distance between the vortex and the antivortex \cite{komineas07}.
Such a VA pair will be called a rotating pair.
As it turns out, the three-vortex process that leads to vortex-polarity
switching involves both types of VA pairs and may be thought of as a
composition of Kelvin and rotational motion.

Thus our aim in this paper is
to study in some detail a three-vortex process in which a Kelvin pair
collides against a single vortex initially at rest. During collision
a transient rotating VA pair forms and its dynamics is crucial for the final
outcome of the scattering process. In short, when the speed of the
incoming Kelvin pair is sufficiently low, the scattering process is
mediated by the formation of a rotating VA pair but is eventually 
semi-elastic; the Kelvin pair reemerges after collision at a peculiar
scattering angle while the initial single vortex comes to rest at a
new position. This unusual behavior can be explained by the
special nature of the conservation laws of linear and angular momentum
due to the underlying nontrivial topological structure.
In contrast, for larger values of the speed of the incoming Kelvin pair,
the intermediate rotating VA pair suffers a topologically forbidden
annihilation (which is possible on a discrete lattice) and thus leads to
a highly inelastic scattering process; the final product is a single
vortex at rest with polarity opposite to the polarity of the original
single vortex.

Numerical simulations are performed on the basis of the 
two-dimensional (2D) Landau-Lifshitz (LL) equation briefly described
in Sec.~II. In Sec.~III we present our main result for semi-elastic
scattering at low velocities, which may be understood as a transmutation
of momentum into position (and vice versa) due to the underlying topology.
Vortex polarity switching occurs at higher velocities and is described
in Sec.~IV. Some concluding remarks are included in Sec.~V. In the 
Appendix we present an analytical solution based on collective coordinates 
which provides a reliable approximation at sufficiently low velocities,
in agreement with the numerical simulations of Sec.~III.  

\section{Magnetic vortices}
\label{sec:vortex}

A ferromagnet is characterized by the magnetization $\bm{m} = (m_1, m_2, m_3)$
measured in units of the constant saturation magnetization $M_s$.
Hence $\bm{m}$ is a vector field of unit length, $\bm{m}^2 = m_1^2+m_2^2+m_3^2 = 1$,
but is otherwise a nontrivial function of position and time
$\bm{m} = \bm{m}(\bm{r},t)$ that satisfies the rationalized Landau-Lifshitz (LL)
equation
\begin{equation}  \label{eq:lle}
\frac{\p\bm{m}}{\p t} = \bm{m} \times \bm{f}, \quad
\bm{f} \equiv \Delta\bm{m} - \qfactor\, m_3\, \bm{\hat{e}}_{\rm 3}, \quad
\bm{m}^2 = 1.
\end{equation}
Here distances are measured in units of the exchange length
$\lex = \sqrt{A/2\pi M_s^2}$, where $A$ is the exchange constant,
and the unit of time is $\tau_0 \equiv 1/(4\pi\gamma M_s)$ where $\gamma$ is the
gyromagnetic ratio.
Typical values are $\lex \sim 5{\rm nm}$
and $\tau_0 \sim  10{\rm ps}$ which set the scales for the phenomena
described by Eq.~(\ref{eq:lle}).
To complete the description of the LL equation we note that we consider
ferromagnetic materials with uniaxial anisotropy.
Then $\bm{\hat{e}}_{\rm 3}$ in Eq.~(\ref{eq:lle}) is a unit vector along
the symmetry axis and the dimensionless parameter $\qfactor \equiv K/2\pi M_s^2$,
where $K$ is an anisotropy constant, measures the strength of anisotropy.
In particular, $\qfactor$ is taken to be positive throughout this paper,
a choice that corresponds to easy-plane ferromagnets.
Also note that we have neglected the demagnetizing field in Eq.~(\ref{eq:lle})
which amounts, in some respects, to a simple additive renormalization
of the anisotropy constant in the thin-film limit \cite{gioia97}.
To be sure, the thin-film limit is more involved \cite{carboux01,kohn05,caputo07}
but some of the key issues can be discussed already within
the simplified model defined by Eq.~(\ref{eq:lle}).
We adopt this limit in the remainder of this paper and thus consider
a two-dimensional (2D) restriction of Eq. (1) while $\qfactor$ is set equal
to unity by a suitable rescaling of the space-time coordinates $x$,$y$ and $t$. 
The effective field $\bm{f}$ in Eq.~(\ref{eq:lle}) may be derived from
a variational argument:
\begin{equation}  \label{eq:energy}
\bm{f} = -\frac{\delta E}{\delta \bm{m}}\,; \quad
E = \half \int{\left[ (\bnabla\bm{m})^2 + m_3^2 \right]\, dx dy}
\end{equation}
where $E$ is the conserved energy functional.

Static solutions are obtained by solving Eq. (1) with the time derivative set equal to zero or, equivalently, by finding stationary points of the energy functional.
We can write the general axially symmetric vortex solution as
\begin{equation}  \label{eq:vortex}
m_1 + i\, m_2 = \sin\theta\, e^{i\kappa(\phi-\phi_0)},
\quad m_3 = \lambda\cos\theta,
\end{equation}
where $\rho$ and $\phi$ are the usual cylindrical coordinates and
$\theta=\theta(\rho)$ is the vortex profile that  can be found
numerically \cite{papanicolaou99}.
The vortex number is $\kappa=+1$ for a vortex and $\kappa=-1$ for an antivortex.
The integer $\lambda=\pm 1$ defines the vortex polarity.
The constant $\phi_0$ is an arbitrary angle that reflects the azimuthal symmetry of Eq. (1) and does not affect the vortex energy.

A key quantity for describing both topological and dynamical properties of the 2D
LL equation is the local topological vorticity $\gamma = \gamma(x,y,t)$ defined
from \cite{papanicolaou91,komineas96}:
\begin{equation}  \label{eq:vorticity1}
\gamma = \half\, \eab\, (\p_\beta\bm{m} \times \p_\alpha\bm{m}) \cdot \bm{m},
\end{equation}
where the usual summation convention is invoked for the repeated indices $\alpha$
and $\beta$, which take over two distinct values corresponding to the two spatial
coordinates $x$ and $y$, and $\eab$ is the 2D antisymmetric tensor.
In particular, one may consider the total topological vorticity $\Gamma$
and the Pontryagin index or skyrmion number $\pontryagin$ defined from
\begin{equation}  \label{eq:skyrmionnumber}
\Gamma = \int{\gamma\, dx dy},\quad  \pontryagin = \frac{\Gamma}{4 \pi}.
\end{equation}
For the vortex given by Eq. ~(\ref{eq:vortex})
the total vorticity $\Gamma$ and the skyrmion number $\pontryagin$
are calculated from Eq.~(\ref{eq:skyrmionnumber}) to be
\begin{equation}  \label{eq:vortextopo}
\Gamma = -2\pi\kappa\lambda\,,\quad \pontryagin = -\half\,\kappa\lambda\,,
\end{equation}
We must thus distinguish four types of vortex states labelled by ($\kappa,\lambda$)
where the vortex number $\kappa=\pm 1$ and the polarity $\lambda=\pm 1$ may be taken
in any combination. In all cases a vortex state is characterized by half-integer
skyrmion number $\pontryagin$, in contrast to magnetic bubbles observed
in easy-axis ferromagnets which carry integer skyrmion number.
Nevertheless, a magnetic vortex shares with a magnetic bubble the fundamental
property that it is spontaneously pinned around a fixed guiding center;
i.e., it cannot move freely unless an external magnetic field gradient
is applied or other vortices are present in its vicinity.

One may also consider a pair of vortices ($\kappa_1,\lambda_1$) and ($\kappa_2,\lambda_2$) with total skyrmion number that is an integer:
\begin{equation}  \label{eq:k1l1k2l2}
           \pontryagin = -\half\,(\kappa_1 \lambda_1 + \kappa_2 \lambda_2).
\end{equation}
The energy of such a configuration is finite if we restrict attention
to vortex-antivortex (VA) pairs ($\kappa_1=-\kappa_2$). For definiteness,
we choose $\kappa_1=1$ and $\kappa_2=-1$ and thus the skyrmion number
\begin{equation}  \label{eq:l1l2}
           \pontryagin = -\half\,(\lambda_1-\lambda_2)
\end{equation}
depends on the polarities $\lambda_1$ and $\lambda_2$. Now, a VA pair with equal 
polarities ($\lambda_1=\lambda_2=\pm 1$) is topologically trivial ($\pontryagin=0$). In contrast,
a VA pair with opposite polarities is topologically equivalent to a skyrmion
($\pontryagin=1$, for $\lambda_1=-1$ and $\lambda_2=1$) or an antiskyrmion
($\pontryagin=-1$, for $\lambda_1=1$ and $\lambda_2=-1$).

Unlike single vortices, the VA pairs sketched above cannot be static solutions
of the LL equation. A topologically trivial $(\pontryagin=0)$ VA pair
undergoes Kelvin motion in a direction perpendicular to the line connecting
the vortex and the antivortex \cite{papanicolaou99}, whereas a topologically
nontrivial ($\pontryagin=\pm 1$) VA pair undergoes a rotating motion around a
fixed guiding center \cite{komineas07}.
The gross features of these two types of motion of a VA pair had
    been anticipated by a treatment based on collective coordinates [9]
    which is valid only in the limit of widely separated vortices.
    In contrast, Refs. [13] and [14] provide a complete field theoretical
    calculation within the LL equation of true solitary waves in Kelvin
    and rotational motion, respectively. Such a calculation remains valid
    for  any relative separation of the VA pair, even when the pair reduces
    to a solitary lump with no apparent topological features and is thus
    beyond the reach of collective coordinates. Both types of motion
    of a VA pair are important for understanding the three-vortex process
    analyzed in the continuation of this paper.

\section{Vortex pair scattering}
\label{sec:scattering}

Recent experiments have shown that a vortex may switch its polarity under
the influence of a very weak magnetic field of the order of a few $mT$
\cite{neudert05,waeyenberge06}. The key to this phenomenon is the appearance of a
vortex-antivortex (VA) 
pair which is spontaneously created in the vicinity of a preexisting
vortex.
We are thus motivated to study in detail this particular system of three vortices.
Here we do not address the question of how a VA pair is actually
created. In fact, production of a VA pair might not be possible in
    the idealized model studied in the present paper.
Rather we concentrate on the second step of the process and explain in
detail how the vortex interacts with the VA pair.

\begin{figure}
\epsfig{file=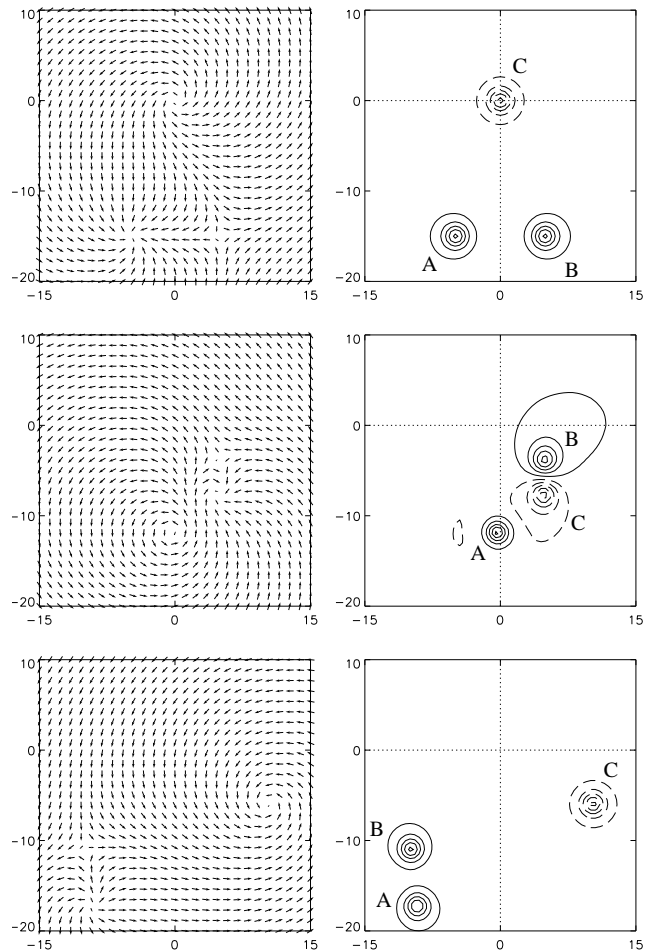,width=1.0\linewidth}
   \caption{Three snapshots for the collision of a VA pair in Kelvin motion
(the AB pair), initially located at $(0,-15)$ and propagating
with velocity $v=0.1$ along the $y$ axis, against a target vortex C initially located
at the origin.
During collision, antivortex B rotates around vortex C before rejoining
its original partner A to form a new VA pair that scatters off at an angle
in the third quadrant.
The target vortex C is shifted to a
new position in the fourth quadrant thanks to transmutation
of VA pair momentum into vortex position.
Left column depicts the 2D projections of the magnetization $(m_1,m_2)$
while the right column depicts the level contours of $m_3$ with solid lines
corresponding to positive values and dashed lines to negative ones.
   }  \label{fig:v01}
\end{figure}

It is obvious that the full dynamics of a three-vortex system
is rather rich.
In order to be able to analyse this system in terms of elementary processes 
we set up the initial configuration as depicted in the first panel of Fig.~\ref{fig:v01}.
Specifically, we assume that a single vortex $(\kappa,\lambda)$=$(1,-1)$=C is initially at
rest at some specified point which is taken to be the origin of the
coordinate system. We further assume that a Kelvin pair (with $\pontryagin = 0$)
consisting of a $(1,1)$=A vortex and a $(-1,1)$=B antivortex is somehow
created in the neighborhood of the original vortex. Once created the AB pair
will undergo Kelvin motion along the $y$ axis and eventually
collide with the single vortex C.

One could use any reasonable VA pair ansatz for the pair AB.
Here, we shall use the VA pairs calculated in Ref.~\cite{papanicolaou99}
as exact steady state solutions of the LL equation.
These are viewed as solitary waves which
propagate with a velocity in the range $0 < v < 1$ where the maximum velocity
$v=1$ coincides with the low-wavenumber
limit of the magnon group velocity expressed in rationalized units.
For small velocities $v \ll 1$ these solitary waves are composed of a vortex A
and an antivortex B at a relative distance  $d \sim 1/v$. 

   The process was simulated by integrating numerically the
corresponding initial-value problem in the LL equation.
Figure \ref{fig:v01} provides an illustration with
three characteristic snapshots in the case of a relatively
slow Kelvin pair initially moving along the $y$ axis with velocity $v=0.1$ for
which the vortex and the antivortex are separated by a distance
$d \approx 1/v=10$. As the pair approaches, the original C=$(1,-1)$ vortex teams
up with the B=$(-1,1)$ partner of the AB pair to form a new, topologically
nontrivial $(\pontryagin =1)$ VA pair in quasi-rotational motion. In fact, B rotates
almost a full circle around C before rejoining its original partner A. The new
AB pair is again a topologically trivial $(\pontryagin =0)$ VA pair in Kelvin motion that
moves away from the target vortex, having suffered a total scattering angle
that is greater than $\pi/2$ from its original direction.
The scattering is slightly
inelastic in the sense that the outgoing AB pair moves out with greater
velocity $(v=0.15)$. And, most remarkably, the target vortex C moves away from
the origin and comes to rest at a new location in the fourth quadrant of the
xy plane.

The unexpected result in this numerical simulation is that the
VA pair AB is scattered at an angle to the y axis which is the initial axis
of propagation. In a naive approach to the problem one would think that 
the VA pair scattering and the corresponding change in its linear momentum
does not seem to be accompanied by a corresponding
change in momentum of any other object taking part in the process.
This would appear to contradict the conservation of linear momentum
in this conservative system.

Such a  peculiar behavior can be explained by the unusual nature of
the linear momentum which allows for a transmutation between
position and impulse in the case of topologically nontrivial systems, such
as the three-vortex system considered here.
The two components of linear momentum (impulse) $\bm{P}=(P_x, P_y)$
can be written as moments of the topological vorticity
$\gamma$ \cite{papanicolaou91,komineas96}:
\begin{equation}  \label{eq:linmom}
P_x = - \int y\gamma\ dx dy, \qquad
P_y = \int x\gamma\ dx dy,
\end{equation}
while the angular momentum is given by
\begin{equation}  \label{eq:angmom}
L = \half \int \rho^2 \gamma\,dx dy
\end{equation}
where $\rho^2=x^2+y^2$.

We first consider the impulse of a single vortex like 
vortex C shown in the first panel of Fig.~\ref{fig:v01} which sits at the origin.
The impulse is clearly zero due to the axial symmetry of the vortex.
It should also be noted that $\bm{P}$ is actually
a measure of position for an isolated vortex.
Indeed, if the vortex is shifted to a new position $(x,y)=(a,b)$,
the corresponding impulse is given by $P_x = b\Gamma$ and $P_y = -a\Gamma$
(with $\Gamma = -2\pi\kappa\lambda$) and clearly provides a measure of position
through the guiding-center coordinates $R_x = -P_y/\Gamma=a$ and $R_y = P_x/\Gamma=b$.
Similarly, the angular momentum $L$ of a static vortex is nonzero because 
the topological vorticity $\gamma$ is of definite sign. $L$ is rather
a measure of vortex size. For a single isolated vortex located at the origin we find numerically 
$L_V = -8.5\kappa\lambda$. 

The linear momentum (impulse) of the AB pair shown in the first panel of Fig.~\ref{fig:v01}
can be calculated approximately if we suppose that the vorticity of each
vortex is concentrated at its center. Considering that the antivortex has
vorticity opposite to that of the vortex, we find $\bm{P}_{AB} \approx (0, 2\pi d)$,
where $d \sim 1/v$ is the relative distance between the vortex and the antivortex.
A more precise calculation for $v=0.1$ yields $P_{AB} = 62$ \cite{papanicolaou99}.

The single vortex C shown in the final panel of Fig.~\ref{fig:v01} is static and
sits in the fourth quadrant at position ($10, -6$).
This corresponds to an impulse $\bm{P} = (2\pi\times 6, 2\pi\times 10)$
according to Eqs.~(\ref{eq:linmom}).
The outgoing AB pair propagates with velocity $v=0.15$
at an angle $\psi=0.64\pi$ with respect to its initial direction of propagation.
The impulse of a pair with $v=0.15$ can be found in the appropriate
table and figure of Ref.~\cite{papanicolaou99} to be $P_{AB}=40$,
or, given the direction of propagation, $\bm{P}_{AB} = (-36,-17)$.
The result is that the AB pair changes its linear momentum by
$\Delta\bm{P}_{AB} = (-36,-56)$ due to scattering.
The corresponding quantity for the single vortex is
$\Delta\bm{P}_C = (2\pi\times 6, 2\pi\times 10) = (38,63)$.
Conservation of linear momentum would require that 
\begin{equation}
\Delta\bm{P}_{AB} + \Delta\bm{P}_C = 0
\end{equation}
in the case of elastic scattering.
This relation is approximately
satisfied in our simulation -- the difference is due to spin-wave radiation
produced during scattering.

The numerical simulation of Fig.~\ref{fig:v01} shows that the linear momentum (impulse)
of a VA pair is transferred to vortex impulse which is tantamount
to a translation in the vortex position. A genuine transmutation
of momentum into position takes place during the three-vortex collision.  
Therefore, the definition of the impulse given in Eq.~(\ref{eq:linmom})
is not only consistent with the symplectic structure of the Landau-Lifshitz
equation \cite{papanicolaou91} but it is indeed physically relevant.

We may now use as the initial VA pair a solitary wave with
larger velocity. We repeat the numerical simulation using a VA pair
with $v=0.2$. The time evolution is similar to that of Fig.~\ref{fig:v01}.
The velocity of the outgoing VA pair is $v=0.46$ and the pair is scattered at
an angle $\psi=0.60\pi$ relative to the initial direction of propagation.
The velocity of the outgoing VA pair is significantly larger that that
of the incoming pair and indicates a significant inelastic component of the process.
The position of the single vortex C after scattering is ($5, -2$).
One can check that the difference in the VA pair linear momentum is
transferred to the single vortex; i.e., to its translated position.

A simple approach to the dynamics of three vortices and specifically
to the phenomena described in this section
can be obtained through a collective collective-coordinate approach 
which is valid for sufficiently low velocity of the incoming VA pair.
This approach is described in detail in the Appendix.
In particular, we are able to deduce theoretically an approximation for the
scattering angle of the VA pair.

\section{Switching of vortex polarity}
\label{sec:switch}

\begin{figure}
\epsfig{file=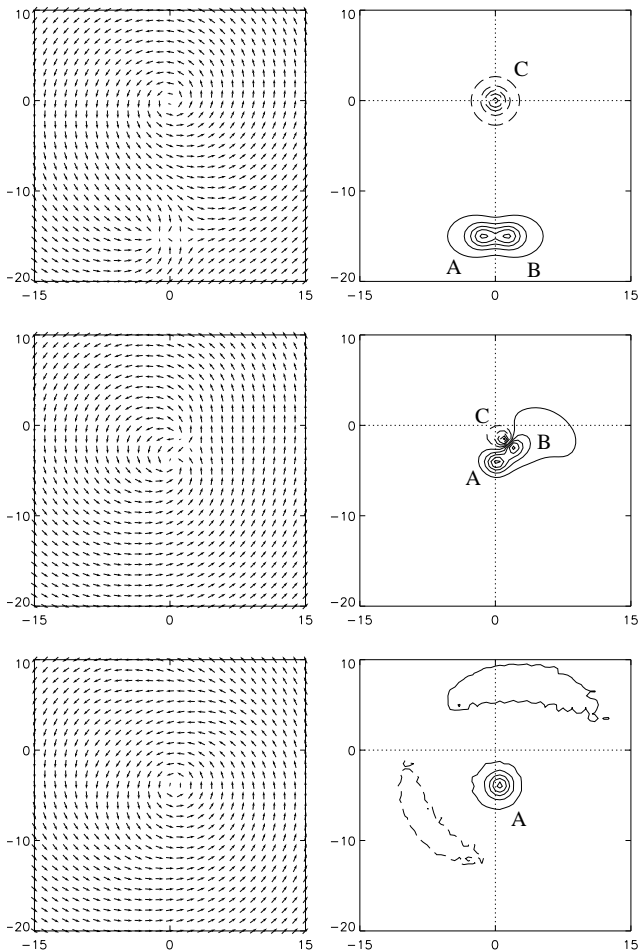,width=1.0\linewidth}
   \caption{
Same as Fig.~\ref{fig:v01} but for a larger initial velocity $v=0.5$ of
the AB pair. During collision, antivortex B begins
to rotate around vortex C but the rotating BC pair is
eventually annihilated leaving behind vortex A (with
polarity opposite to that of the target vortex C) and
a burst of spin waves that propagate away from the scattering region.
   }  \label{fig:v05}
\end{figure}

   The preceding numerical experiment was repeated for a Kelvin pair
with relatively large velocity $v=0.5$ for which the vortex and the antivortex
are tightly bound at a relative distance $d=2.6$ \cite{papanicolaou99}. The process is
again illustrated by three characteristic snapshots in Figure \ref{fig:v05}.
While the initial stages of the process are similar to those encountered in the case
of slow Kelvin motion (Figure \ref{fig:v01}) a substantial departure occurs when the
pair now approaches the target vortex. In particular, as soon as antivortex
B=$(-1,1)$ begins to rotate around the target vortex C=$(1,-1)$ they collide and
undergo a spectacular $\Delta\pontryagin =1$ transition (annihilation) leaving behind the
A=$(1,1)$ vortex which may be thought of as the target vortex C with
polarity flipped from $-1$ to 1 (vortex core switching) and a burst of spin
waves propagating away from the scattering region.

A crucial element in the preceding numerical experiment is the transient
formation of a rotating VA pair (the BC pair) which is characterized by a
nonzero skyrmion number $\pontryagin=1$. Nevertheless, the BC pair is
annihilated during collision and thus leads to a topologically forbidden
transition that changes the skyrmion number by one unit or, equivalently,
the polarity of the original vortex is flipped from $-1$ to $+1$. 
Such a process is indeed possible because a topologically nontrivial
(and thus rotating) VA pair may shrink without encountering an energy
barrier and eventually be annihilated when its size approaches the
spacing of the underlying lattice. An excess amount of energy equal to $4\pi$ is then released into the system in the form of spin waves \cite{komineas07}. 

\begin{figure}
\epsfig{file=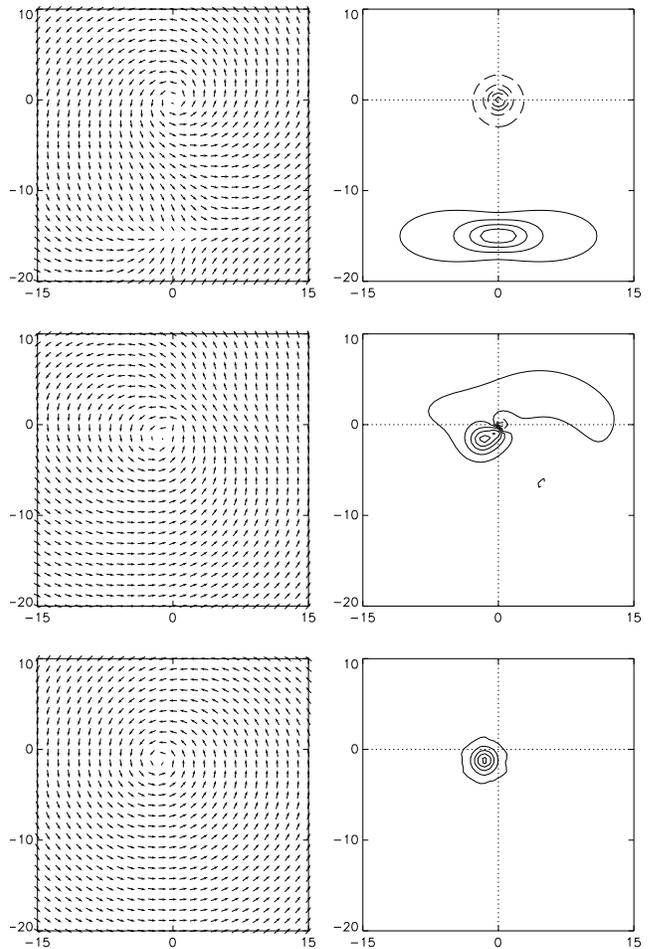,width=1.0\linewidth}
   \caption{
Same as Fig.~\ref{fig:v01} but for a solitary wave with initial velocity $v=0.9$.
During collision the solitary wave is transformed into a VA pair.
After collision the initial vortex has been annihilated with the generated antivortex
while the generated vortex remains static near the origin.
The net result may again be thought as the original (target) vortex with polarity flipped from $-1$ to $+1$.
   }  \label{fig:v09}
\end{figure}

\begin{figure}
\epsfig{file=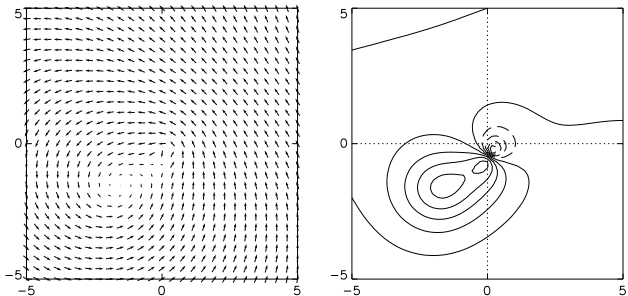,width=1.0\linewidth}
   \caption{The second snapshot of Fig.~\ref{fig:v09} magnified.
The VA pair created out of the incoming solitary wave is clearly shown here.
The generated antivortex is about to be annihilated together with the original (target) vortex.
   }  \label{fig:v09_blowup}
\end{figure}

According to the results in Ref.~\cite{papanicolaou99} solitary waves
with velocities $0.78 < v < 1$ apparently lose their vortex-antivortex character.  Rather they are droplets of  nonuniform magnetization. In particular, the magnetization does not acquire the values $\bm{m}=(0,0,\pm 1)$
at any point, thus no vortex centers can be identified.
In the remainder of this section we shall present  simulations of the
interaction of such solitary waves with a target vortex.
We first choose an incoming solitary wave with velocity $v=0.9$ which may indeed be viewed as a weak disturbance of the ferromagnetic vacuum, as shown in the first panel of
Fig.~\ref{fig:v09}.
As the solitary wave collides with the target vortex at the origin, it is
transformed into a vortex and an antivortex, as seen in the second panel of
Fig.~\ref{fig:v09}. The same snapshot is shown magnified in Fig.~\ref{fig:v09_blowup}.
The antivortex is then annihilated with the target vortex
and finally a vortex remains with polarity opposite to the initial one.

\begin{figure}
\epsfig{file=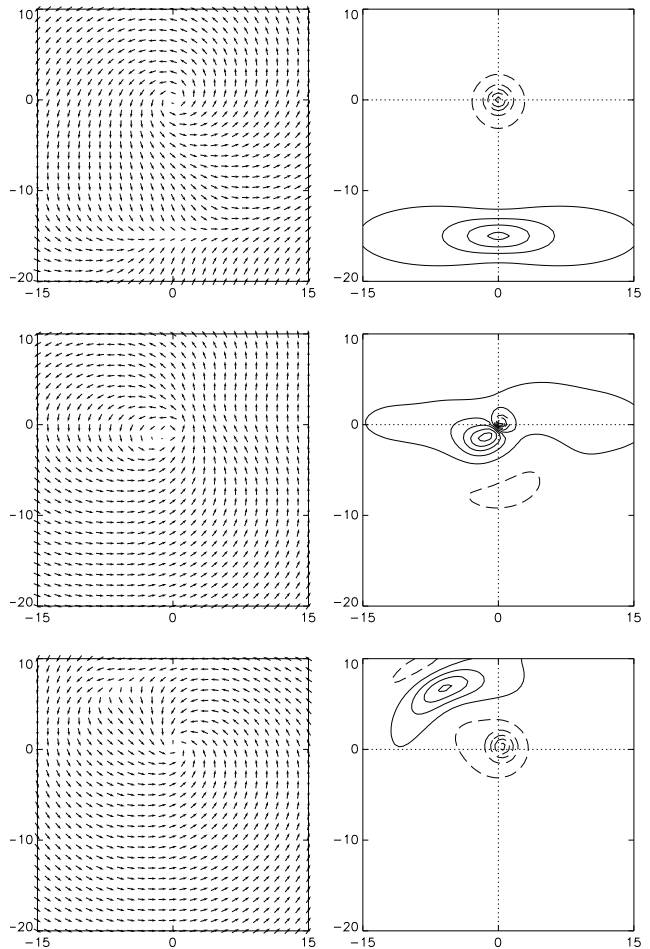,width=1.0\linewidth}
   \caption{Three snapshots for the collision of a solitary wave
initially located at $(0,-15)$ and propagating
with velocity $v=0.95$, against a target vortex initially located
at the origin.
During collision, the solitary wave as well as the vortex shrink,
and finally the solitary wave is deflected in the second quadrant.
The target vortex remains unaffected except that it is shifted slightly to a
new position in the first quadrant thanks to transmutation
of solitary wave momentum into vortex position.
   }  \label{fig:v095}
\end{figure}

Finally, we present a simulation of the scattering
of a very fast solitary wave with velocity $v=0.95$ which is close to the magnon velocity $v=1$. Again, this wave does not display 
an apparent vortex-antivortex character, as can be seen in the
first panel of Fig.~\ref{fig:v095}.
The solitary wave appears to shrink as it approaches the vortex at the origin
and is deflected into the second quadrant.
The deflected solitary wave is apparently not identical to the initial one due
to energy radiation during scattering.
The small deflection angle indicates a small change of the linear momentum
during scattering, even though this is difficult to quantify accurately from the numerical data. But the
observed  small translation of the vortex from the origin after scattering
indicates a correspondingly small exchange of impulse between the solitary wave and the vortex.
No polarity switching occurs in this case.

To summarize,
a detailed numerical investigation of the three-vortex process for
Kelvin waves with velocities in the allowed range $0<v<1$ suggests the
existence of three characteristic regions separated by two critical
velocities $v_1=0.3$ and $v_2=0.9$ such that $0<v_1<v_2<1$.
For $0<v<v_1$, the Kelvin pair
undergoes nearly elastic scattering of the type depicted in Figure \ref{fig:v01}.
For $v_1<v<v_2$, the process leads to a topologically forbidden $\Delta\pontryagin =1$
transition of the type illustrated in Figure \ref{fig:v05} and Figure \ref{fig:v09}.
There is also some evidence that fast
Kelvin waves with velocities in the narrow range $v_2<v<1$ undergo a nearly
elastic scattering without inversion of the polarity of the target vortex, as illustrated in Figure \ref{fig:v095}.

\section{Concluding remarks}
\label{sec:conclusions}

A direct link between topology and dynamics has been known to exist in 
ferromagnetic media. In particular,
the conservation laws of linear and angular momenta are profoundly
affected by the underlying topology and explain some of the unusual dynamical
features of topological magnetic solitons \cite{papanicolaou91,komineas96}.
For example, a single magnetic vortex cannot move freely but is always
spontaneously pinned around a fixed guiding center. In contrast,
VA pairs with zero skyrmion number ($\pontryagin=0$) can move
rigidly with constant velocity, in analogy with the
Kelvin motion encountered in fluid dynamics
\cite{papanicolaou99}. Free translational motion is again inhibited
for VA pairs with nonzero skyrmion number $\pontryagin$. Instead,
such pairs undergo rotational motion around a fixed guiding center
\cite{komineas07}.

In the present paper we have taken the argument much further.
Specifically, we have studied a three-body collision during which
a VA pair in Kelvin motion is scattered off an isolated single vortex.
The analysis provided interesting results at low velocities of
the incoming Kelvin pair where the process is semi-elastic. 
The total skyrmion number is then preserved and the outcome
of the collision is again a Kelvin pair and an isolated vortex. But the
overall configuration after collision appears to be rather strange
from the point of view of ordinary scattering theory. Specifically,
the outgoing pair is scattered at a peculiar angle while the isolated
vortex comes to rest at a different location. These facts would appear
to violate conservation of linear momentum. However, the observed
transmutation of momentum into position is fully consistent with 
the conservation laws quoted in Sec.~\ref{sec:scattering} and is indeed
due to their special topological structure. Also note that an elegant
mechanical analog of the three-vortex collision is provided by the
analytical solution in terms of collective coordinates given in the
Appendix. Finally, it is now evident that momentum-position transmutation
is a generic and physically relevant feature of multivortex collisions
and is thus the main new result of this paper. 

The process of three-vortex collision
is also thought to be responsible for the vortex
polarity switching observed in ferromagnetic elements
\cite{neudert05,waeyenberge06,yamada07} which entails a change
of the skyrmion number by one unit. A mechanism that changes
the topological number of a magnetic configuration makes it
possible to obtain a controlled switching between topologically
distinct (and thus robust) magnetic states, an issue of obvious
interest for practical applications.
Indeed, our numerical simulations within the 2D Landau-Lifshitz 
equation confirm the possibility of polarity switching when
an incoming VA pair in Kelvin motion with sufficiently high velocity
collides with an isolated vortex. The process is highly inelastic
in that a transient rotating VA pair is formed and subsequently
annihilated, leading to a change of the skyrmion number by one 
unit. This unusual phenomenon is due to the fact that a
topologically nontrivial (and thus rotating) VA pair can shrink
during collision without encountering an energy barrier and be
annihilated when its size approaches the lattice spacing \cite{komineas07}.

\begin{acknowledgments}
N.P. is grateful for hospitality at the Max-Planck-Institute for the Physics
of Complex Systems (Dresden) during completion of this work.
\end{acknowledgments}

\begin{appendix}
\section{Collective coordinates}

\begin{figure}
\epsfig{file=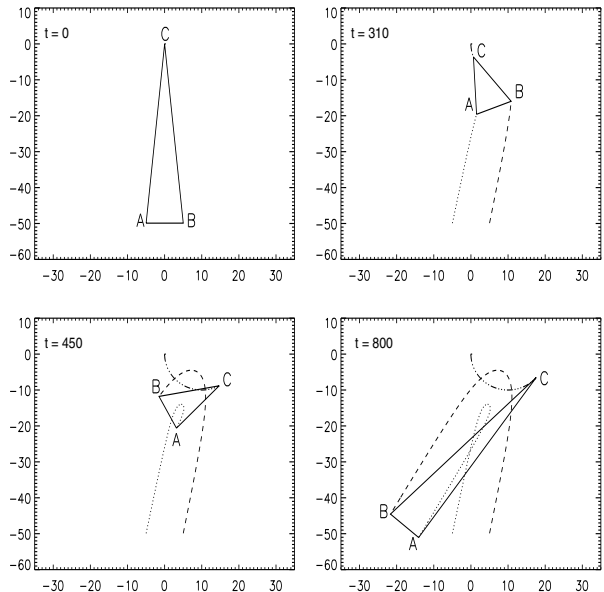,width=1.0\linewidth}
   \caption{Four characteristic snapshots of the three-vortex collision calculated
within a collective-coordinate approach starting from an initial configuration
with $h_0=50$ and $d=10$. The three (anti)vortices form an isosceles triangle (AC=BC)
with base of constant length AB=d at all times. See the Appendix for further explanations. 
   }  \label{fig:collective}
\end{figure}

The main results of Sec.~\ref{sec:scattering} can be understood within
a formulation in terms of collective coordinates where the three vortices A=$(1,1)$, B=$(-1,1)$ and C=$(1,-1)$  are approximated as pointlike particles whose coordinates 
$\bm{R}_i = (X_i, Y_i)$, with $i=1,2,3$, satisfy the equations of motion \cite{thiele74,pokrovskii85,kovalev02}
\begin{equation}  \label{eq:collectiveEq1}
2\pi\kappa_i\lambda_i\, \left(\frac{d\bm{R}_i}{dt} \times \bm{\hat{z}}\right) =
2\pi\kappa_i\, \sum_{j\neq i} \kappa_j\, \frac{\bm{R}_i-\bm{R}_j}{|\bm{R}_i-\bm{R}_j|^2}.
\end{equation}
We note that the mutual forces on the rhs depend only on the vortex numbers ($\kappa$)
and are attractive for pairs AB and BC and repulsive for AC.
In contrast, the lhs of Eq.~(\ref{eq:collectiveEq1})
depends on the vortex skyrmion number and
thus on both the vortex number ($\kappa$) and the polarity ($\lambda$).
A more explicit form of Eq.~(\ref{eq:collectiveEq1}) is given by
\begin{equation}  \label{eq:collectiveEq2}
\lambda_i\, \left(\frac{dY_i}{dt}, -\frac{d X_i}{dt}\right) =
 \sum_{j\neq i} \kappa_j\, \frac{\bm{R}_i-\bm{R}_j}{|\bm{R}_i-\bm{R}_j|^2},
\end{equation}
applied for i=1,2 and 3,  which leads to a system of six first-order differential equations that must be solved with initial condition
\begin{eqnarray}  \label{eq:initialconfig}
   X_1 & = & -d/2, \quad   X_2 = d/2, \quad   X_3 = 0  \nonumber \\
   Y_1 & = & -h_0, \quad\;\;   Y_2 = -h_0,\quad   Y_3 = 0
\end{eqnarray}
which corresponds to the scattering process studied in Sec.~\ref{sec:scattering}.
In words, the topologically trivial ($\pontryagin=0$) AB pair begins to move
from a distance $h_0$ along the negative $y$ axis with velocity $v=1/d$ against a vortex C
initially located at the origin.
The initial configuration of the three vortices A, B and C is shown in the first panel
of Fig.~\ref{fig:collective} which makes it clear that the triangle ABC
is initially isosceles (AC=BC).

The equations of motion (\ref{eq:collectiveEq1}) and (\ref{eq:collectiveEq2})
lead to four conservation laws; namely, the energy
\begin{equation}  \label{eq:collectiveEnergy}
E = 2\pi\, \ln \left[\frac{|\bm{R}_1-\bm{R}_2|\,|\bm{R}_2-\bm{R}_3|}{|\bm{R}_1-\bm{R}_3|}\right]
\end{equation}
the two components of linear momentum (impulse)
\begin{eqnarray}  \label{eq:collectiveLinmom}
P_x & = & 2\pi\, (Y_1 - Y_2 - Y_3) \nonumber \\
P_y & = & -2\pi\, (X_1 - X_2 - X_3).
\end{eqnarray}
and the angular momentum
\begin{equation}  \label{eq:collectiveAngmom}
L =\pi\, \left[-(X_1^2+Y_1^2) + (X_2^2+Y_2^2) + (X_3^2+Y_3^2) \right].
\end{equation}
The actual values of the preceding  conserved quantities may be calculated
from the initial configuration defined in Eq.~(\ref{eq:initialconfig}), namely
\begin{equation}  \label{eq:conservedValues}
E=2\pi\,\ln(d),\quad P_x=0,\quad P_y=2\pi d,\quad L=0.
\end{equation}
However  the above four conservation laws are not sufficient to readily solve
the six differential equations contained in Eq. (\ref{eq:collectiveEq2}).

Hence, we first solved the equations of motion (\ref{eq:collectiveEq2})
numerically with initial condition given by Eq.~(\ref{eq:initialconfig}).
The main features of the solution can be summarized as follows:

1. The triangle ABC remains isosceles (AC=BC) at all times,
while the length of the base AB remains constant. Thus we may write
\begin{equation}
       AB = d,\qquad     AC = BC = \sqrt{h^2+\frac{d^2}{4}}
\end{equation}
where only the height of the triangle depends on time, $h=h(t)$,
and satisfies the initial condition $h(t=0)=h_0$.

2. There is a characteristic time $t_0$ at which $h(t=t_0)=0$ and the triangle
degenerates into a straight line element ACB with vortex C located
at a common distance d/2 from both A and B. Therefore, $t_0$ is the instance of closest
encounter of all three (anti)vortices.

3. For $t>t_0$, C emerges from the other side and forms again an isosceles triangle ABC.
In the meantime, antivortex B encircles vortex C while keeping a constant distance d
from its Kelvin partner A. Eventually, a new Kelvin pair AB is
formed which moves off to infinity at a scattering angle in the third quadrant with asymptotic
velocity equal to $v=1/d$.   

4. Throughout the process vortex C moves on the lower semicircle of the circle 
\begin{equation}
            (X_3-d)^2 + Y_3^2 = d^2
\end{equation}
In fact, it covers only a portion of the lower semicircle, up to a maximum
angle that depends on $h_0$ and $d$.

The preceding observations are consistent with the conservation laws and suggest the following parametrization of the six coordinates:
\begin{eqnarray}  \label{eq:parametrization}
X_1 & = &  d -\frac{3 d}{2}\,\cos\psi + h \sin\psi, \;\;
Y_1 = -\frac{3 d}{2}\,\sin\psi - h\cos\psi   \nonumber  \\
X_2 & = &  d -\frac{d}{2}\,\cos\psi + h \sin\psi, \quad
Y_2 = -\frac{d}{2}\,\sin\psi - h\cos\psi     \nonumber \\
X_3 & = &  d\,(1-\cos\psi), \quad
Y_3 = -d\,\sin\psi
\end{eqnarray}
in terms of only two variables; namely, the height of the triangle $h=h(t)$
and the scattering angle $\psi=\psi(t)$ which is the angle between the symmetry axis
of the triangle and the $y$ axis. 
Then all six equations in (\ref{eq:collectiveEq2}) reduce to the pair of equations
\begin{eqnarray}  \label{eq:collectiveEq3}
\dot{\psi} = \frac{1}{\Delta^2},\quad & & \dot{h}=-\frac{1}{d}\,\frac{\Delta^2 + d^2}{\Delta^2},
   \\
 \Delta^2 & \equiv & h^2 + \frac{d^2}{4}, \nonumber
\end{eqnarray}
where the overdot denotes time derivative, which must be solved with initial condition $h(t=0)=h_0$ and $\psi(t=0)=0$.

The second equation in (\ref{eq:collectiveEq3}) can be integrated to yield
\begin{eqnarray}  \label{eq:h}
t = t_0 - \gamma d^2\,
  \left[\frac{h}{\gamma d} - \frac{4}{5}\, \arctan\left(\frac{h}{\gamma d}\right) \right],
\nonumber \\
\gamma \equiv \frac{\sqrt{5}}{2},\quad
t_0\equiv \gamma d^2\,
  \left[\frac{h_0}{\gamma d} - \frac{4}{5}\, \arctan\left(\frac{h_0}{\gamma d}\right) \right].
\end{eqnarray}
Then we combine both equations in (\ref{eq:collectiveEq3}) to write
$dh/d\psi = -(1/d)(\Delta^2+d^2)$ which can also be integrated to yield
\begin{equation}  \label{eq:psi}
\frac{h}{\gamma d} = \tan[\gamma (\psi_0-\psi)],\quad
\frac{h_0}{\gamma d} \equiv \tan[\gamma \psi_0].
\end{equation}
In words, at $t=0,\, h=h_0$ and $\psi=0$. At $t=t_0$, $h$ vanishes and $\psi$ reaches
the value $\psi_0$. For $t > t_0$, $h$ reemerges on the other side and assumes
negative values in our conventions.
In the far future ($t \to \infty$), $h \sim -t/d \to -\infty$ and the scattering
angle reaches the maximum value defined from $\gamma (\psi_0 - \psi_{\rm max})=-\pi/2$
or
\begin{equation}  \label{eq:psimax}
\psi_{\rm max} = \psi_0 + \frac{\pi}{2\gamma}.
\end{equation}
In particular, if $h_0 \to \infty$, $\psi_0 \to \pi/(2\gamma)$ and
\begin{equation}  \label{eq:psimax1}
\psi_{\rm max}=\frac{\pi}{\gamma} = \frac{2\pi}{\sqrt{5}}, \quad ({\rm as}\;\; h\to \infty).
\end{equation}
The preceding results are partially illustrated in Fig. ~\ref{fig:collective}
for an initial configuration specified by $h_0=50$, $d=10$ and $\psi=0$.

In order to compare with the results on vortex scattering presented in Fig.~\ref{fig:v01}
we choose $h_0=15$, $d=9.5$ and calculate $t_0=65.4$, $\psi_0=0.854$ and
$\psi_{\rm max} = 0.72\pi$. In the case of the  simulation shown in Fig.~\ref{fig:v01}
we have found $\psi_{\rm max} = 0.64\pi$.
This number as well as the overall picture are in good agreement with
the results obtained by the collective coordinate approach.
The observed small differences should be mainly due to the radiation present
in the actual simulations and are expected to increase at larger velocity
of the incoming Kelvin (AB) pair. Indeed, while agreement with the
collective-coordinate approach persists for $v$=0.2, the picture changes
drastically for higher velocities, as  discussed in Sec.~\ref{sec:switch}.

\end{appendix}


\begin{thebibliography}{10}

\bibitem{hubert}
A. Hubert and R. Sch\"afer, {\em Magnetic domains} (Springer, Berlin, 1998).

\bibitem{huber82}
D.~L. Huber, Phys. Rev. B {\bf 26},  3758  (1982).

\bibitem{mertens99}
F. Mertens and A. Bishop, ``Dynamics of Vortices in Two-Dimensional Magnets''
  in P.L. Christiansen and M.P. Sorensen (eds.), ``Nonlinear Science at the
  Dawn of the 21st Century'', Springer, 1999.

\bibitem{raabe00}
J. Raabe, R. Pulwey, R. Sattler, T. Schweib\"ock, J. Zweck, and D. Weiss,
J. Appl. Phys. {\bf 88},  4437  (2000).

\bibitem{shinjo00}
T. Shinjo, T. Okuno, R. Hassdorf, K. Shigeto, and T. Ono,
Science {\bf 289},  930  (2000).

\bibitem{wachowiak02}
A. Wachowiak, J. Wiebe, M. Bode, O. Pietzsch, M Morgenstern, and R. Wiesendanger,
Science {\bf 298}, 577 (2002).

\bibitem{shigeto02}
K. Shigeto, T. Okuno, K. Mibu, T. Shinjo, and T. Ono,
Appl. Phys. Lett. {\bf 80},  4190  (2002).

\bibitem{castano03}
F.~J. Castano, C.~A. Ross, C. Frandsen, A. Eilez, D. Gil, H.~I. Smith, M. Redjdal, and F.~B. Humphrey,
Phys. Rev. B {\bf 67},  184425  (2003).

\bibitem{pokrovskii85}
V.~L. Pokrovskii and G.~V. Uimin, JETP Lett. {\bf 41},  128  (1985).

\bibitem{neudert05}
A. Neudert, J. McCord, R. Sch\"afer, and L. Schultz, JAP {\bf 97},  10E701
  (2005).

\bibitem{waeyenberge06}
B.~V. Waeyenberge, A. Puzic, H. Stoll, K.~W. Chou, T. Tyliszczak, R. Hertel, M. F\"ahnle, H. Br\"uckl, K. Rott, G. Reiss, I. Neudecker, D. Weiss, C.~H. Back, and G. Sch\"utz,
Nature(London) {\bf 444},  461  (2006).

\bibitem{yamada07}
K. Yamada, S. Kasai, Y. Nakatani, K. Kobayashi, H. Kohno, A. Thiaville, and T. Ono,
Nature Materials {\bf 6},  269  (2007).

\bibitem{papanicolaou99}
N. Papanicolaou and P.~N. Spathis, Nonlinearity {\bf 12},  285  (1999).

\bibitem{komineas07}
S. Komineas, Phys. Rev. Lett. {\bf 99},  117202  (2007).

\bibitem{gioia97}
G. Gioia and R.~D. James, Proc. R. Soc. Lond. A {\bf 453},  213  (1997).

\bibitem{carboux01} G. Carboux, Math. Mod. Meth. Appl. Sci. {\bf 11}, 1529 (2001).

\bibitem{kohn05} R. Kohn and V.V. Slastikov, Arch. Rat. Mech. Anal. {\bf 178}, 227 (2005).

\bibitem{caputo07} J.~G. Caputo, Y. Gaididei, V.~P. Kravchuk, F.~G. Mertens and
         D.~D. Sheka, Phys. Rev. B {\bf 76}, 174428 (2007).

\bibitem{papanicolaou91}
N. Papanicolaou and T.~N. Tomaras, Nucl. Phys. B {\bf 360},  425  (1991).

\bibitem{komineas96}
S. Komineas and N. Papanicolaou, Physica D {\bf 99},  81  (1996).

\bibitem{thiele74}
A.~A. Thiele, J. Appl. Phys. {\bf 45},  377  (1974).

\bibitem{kovalev02}
A. Kovalev, S. Komineas, and F.~G. Mertens, Eur. Phys. J. B {\bf 65},  89
  (2002).

\end{thebibliography}
\end{document}